\documentclass[twocolumn]{article}
\usepackage[T1]{fontenc}
\usepackage[utf8]{inputenc}
\usepackage{times}
\usepackage{amsmath}
\usepackage{amssymb}
\usepackage{url}

\usepackage{tikz}
\usetikzlibrary{arrows.meta, positioning, shapes.geometric, fit, decorations.pathreplacing}

\usepackage{listings}
\usepackage{xcolor} 

\usepackage{verbatim}
\usepackage{algorithm}
\usepackage{algpseudocode}
\usepackage{graphicx}
\usepackage{booktabs}

\usepackage{authblk}
\usepackage[numbers]{natbib}
\begin{document}
\title{Benchmarking Embedding Models for ESG Data\thanks{This is the author's version of the paper published in the proceedings of ITADATA2026: The 5th Italian Conference on Big Data and Data Science, November 10--12, 2026, Bari, Italy.}}

\author[1]{Motaz Saad\thanks{\texttt{motazk.saad@unisalento.it}, ORCID: 0000-0002-1080-7276}}
\author[1]{Veronica Cret\`i\thanks{\texttt{veronica.creti@unisalento.it}}}
\author[2]{Ivan Gentile\thanks{\texttt{ivan.gentile@ifabfoundation.org}}}
\author[2]{Kianna Kazemi\thanks{\texttt{kianna.kazemi@ifabfoundation.org}}}
\author[1]{Antonella Longo\thanks{\texttt{antonella.longo@unisalento.it}, ORCID: 0000-0002-6902-0160}}
\affil[1]{University of Salento, Lecce, 73100, Italy}
\affil[2]{IFAB Foundation, Bologna, Italy}

\date{}

\maketitle

\begin{abstract}
The use of Environmental, Social, and Governance (ESG) data is fundamental for modern corporate accountability, sustainability reporting, and financial decision-making. Embedding models have emerged as a powerful approach for transforming unstructured ESG text into numerical representations suitable for downstream natural language processing (NLP) tasks. However, their effectiveness in these ESG-specific tasks has not been systematically studied. In this paper, we construct a benchmark dataset specifically tailored to the ESG domain. We benchmark fourteen models, both open-source and closed-source embedding models, comparing their performance with respect to retrieval, and Retrieval-Augmented Generation (RAG). The results demonstrate performance variations across different models, with Qwen3-based models achieving the highest overall performance. This study provides practical insights into which models are better suited for ESG RAG tasks.
\end{abstract}


\section{Introduction}

Environmental, Social, and Governance (ESG) data has become a cornerstone for decision-making in finance, sustainability reporting, and corporate accountability. Organizations, regulators, and investors increasingly rely on ESG information to assess risks, identify opportunities, and evaluate corporate performance beyond financial metrics \cite{rusu2024sustainability, orsolin2024esg}. However, ESG data is inherently heterogeneous: it originates from diverse sources such as sustainability reports, regulatory filings, news articles, and social media. This diversity creates challenges in extracting, representing, and comparing information across contexts.

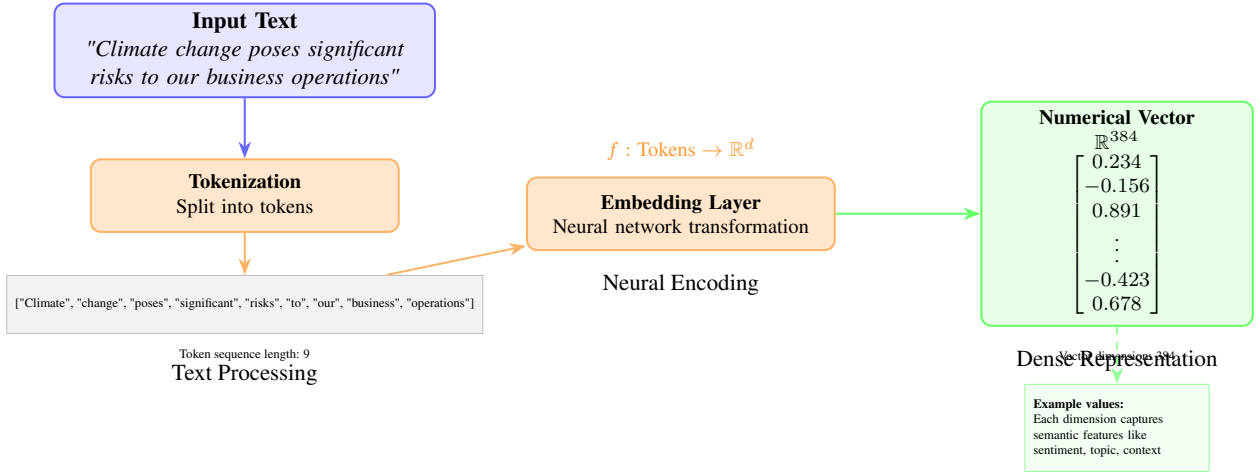
\begin{figure*}
\centering
\resizebox{\textwidth}{!}{%
\begin{tikzpicture}[
    text_input/.style={rectangle, rounded corners, draw=blue!60, fill=blue!10, thick, minimum width=3cm, minimum height=1cm, text centered, font=\small},
    process_step/.style={rectangle, rounded corners, draw=orange!60, fill=orange!20, thick, minimum width=3cm, minimum height=1cm, text centered, font=\footnotesize},
    vector_output/.style={rectangle, rounded corners, draw=green!60, fill=green!10, thick, minimum width=2.5cm, minimum height=2cm, text centered, font=\footnotesize},
    arrow/.style={-Stealth, thick},
    dimension/.style={font=\tiny, text=gray!70}
]

\node[text_input, text width=5cm] (input_text) at (0, 6) {
    \textbf{Input Text}\\
    \textit{"Climate change poses significant risks to our business operations"}
};

\node[process_step, text width=4cm] (tokenization) at (0, 4) {
    \textbf{Tokenization}\\
    Split into tokens
};

\node[rectangle, draw=gray!50, fill=gray!10, minimum width=4.5cm, minimum height=0.8cm, font=\tiny] (tokens) at (0, 2.5) {
    ["Climate", "change", "poses", "significant", "risks", "to", "our", "business", "operations"]
};

\node[process_step, text width=4cm] (embedding) at (6, 3.75) {
    \textbf{Embedding Layer}\\
    Neural network transformation
};

\node[vector_output, text width=3.5cm] (output_vector) at (12, 3.75) {
    \textbf{Numerical Vector}\\
    $\mathbb{R}^{384}$\\
    $\begin{bmatrix}
    0.234 \\
    -0.156 \\
    0.891 \\
    \vdots \\
    -0.423 \\
    0.678
    \end{bmatrix}$
};

\draw[arrow, blue!60] (input_text) -- (tokenization);
\draw[arrow, orange!60] (tokenization) -- (tokens);
\draw[arrow, orange!60] (tokens) -- (embedding);
\draw[arrow, green!60] (embedding) -- (output_vector);



\node[dimension, text=gray!200] at (0, 1.8) {Token sequence length: 9};
\node[dimension, text=gray!200] at (12, 1.8) {Vector dimension: 384};

\node[rectangle, draw=green!40, fill=green!5, minimum width=2.5cm, minimum height=1.2cm, font=\tiny, text width=2.3cm] (values_detail) at (12, 0.8) {
    \textbf{Example values:}\\
    Each dimension captures\\
    semantic features like\\
    sentiment, topic, context
};

\draw[arrow, dashed, green!40] (output_vector) -- (values_detail);

\node[above=0.3cm of input_text, font=\large\bfseries, text=black] {Text Vectorization Process};

\node[below=0.3cm of tokens, font=\small, text=gray!200] {Text Processing};
\node[below=0.2cm of embedding, font=\small, text=gray!200] {Neural Encoding};
\node[below=0.2cm of output_vector, font=\small, text=gray!200] {Dense Representation};

\node[above=0.1cm of embedding, font=\footnotesize, text=orange!80] {$f: \text{Tokens} \rightarrow \mathbb{R}^d$};


\end{tikzpicture}%
}
\caption{Text Vectorization Process}
\label{fig:text_vectorization}
\end{figure*}

Text vectorization refers to the process of transforming unstructured text into numerical representations that can be processed by machine learning models \cite{NLPBook25} as shown in Figure \ref{fig:text_vectorization}. Since computers operate on numbers rather than words, vectorization bridges the gap between natural language and computational models. Early approaches include bag-of-words (BoW) and term frequency–inverse document frequency (TF–IDF), which represent text based on word occurrence counts but ignore word order and semantic meaning \cite{NLPBook25}. More advanced approaches rely on dense vector embeddings, where each text unit (word, sentence, or document) is mapped to a point in a continuous vector space that captures semantic relationships \cite{NLPBook25}.

Word embeddings are a specific type of dense vector representation where individual words are projected into a continuous space such that words with similar meanings or usage contexts are located close to each other \cite{NLPBook25}. Unlike sparse representations such as BoW, embeddings capture semantic similarity (e.g., “climate” and “sustainability” appearing close together) and syntactic patterns (e.g., “company” and “organization”). Popular methods include Word2Vec \cite{word2vec1,word2vec2}, GloVe \cite{glove1,glove2}, and FastText \cite{FastText1,FastText2}, while more recent transformer-based models (e.g., BERT \cite{BERT, transformer}, sentence transformers) extend embeddings beyond words to represent phrases, sentences, and documents.

\begin{figure*}
\centering
\resizebox{\textwidth}{!}{%
\begin{tikzpicture}[
    text_box/.style={rectangle, rounded corners, draw=blue!60, fill=blue!10, thick, minimum width=3cm, minimum height=1.2cm, text centered, font=\footnotesize},
    embedding/.style={rectangle, rounded corners, draw=green!60, fill=green!10, thick, minimum width=2.5cm, minimum height=0.8cm, text centered, font=\footnotesize},
    task/.style={rectangle, rounded corners, draw=purple!60, fill=purple!10, thick, minimum width=2.8cm, minimum height=1cm, text centered, font=\tiny},
    arrow/.style={-Stealth, thick},
    process/.style={rectangle, rounded corners, draw=orange!60, fill=orange!10, thick, minimum width=2.2cm, minimum height=0.8cm, text centered, font=\footnotesize}
]

\node[text_box] (esg_reports) at (0, 8) {ESG Reports};
\node[text_box] (policies) at (0, 6.5) {Company Policies};
\node[text_box] (news) at (0, 5) {News Articles};
\node[text_box] (regulations) at (0, 3.5) {Regulatory Documents};

\node[process] (embedding_model) at (5, 5.75) {Embedding Model};

\node[embedding] (vectors) at (9, 5.75) {Numerical Vectors $\mathbb{R}^d$};

\node[task] (classification) at (13, 8.5) {Document Classification (ESG topics)};
\node[task] (similarity) at (13, 7) {Semantic Similarity \& Clustering};
\node[task] (retrieval) at (13, 5.5) {Information Retrieval \& QA (RAG)};
\node[task] (entity) at (13, 4) {Entity Recognition \& Linking};
\node[task] (sentiment) at (13, 2.5) {Sentiment \& Stance Detection};

\draw[arrow, blue!60] (esg_reports) -- (embedding_model);
\draw[arrow, blue!60] (policies) -- (embedding_model);
\draw[arrow, blue!60] (news) -- (embedding_model);
\draw[arrow, blue!60] (regulations) -- (embedding_model);

\draw[arrow, green!60] (embedding_model) -- (vectors);

\draw[arrow, purple!60] (vectors) -- (classification);
\draw[arrow, purple!60] (vectors) -- (similarity);
\draw[arrow, purple!60] (vectors) -- (retrieval);
\draw[arrow, purple!60] (vectors) -- (entity);
\draw[arrow, purple!60] (vectors) -- (sentiment);

\node[above=0.2cm of esg_reports, font=\small\bfseries, text=blue!80] {Unstructured ESG Text};
\node[above=0.2cm of embedding_model, font=\small\bfseries, text=orange!80] {Transformation};
\node[above=0.2cm of vectors, font=\small\bfseries, text=green!80] {Numerical Representations};
\node[above=0.2cm of classification, font=\small\bfseries, text=purple!80] {Downstream NLP Tasks};

\draw[decorate, decoration={brace, amplitude=5pt}, thick, blue!60] 
    (-1.5, 2.5) -- (-1.5, 8.8);
\node[left=0.1cm of esg_reports, rotate=90, font=\small, text=blue!80] at (-2, 5.75) {ESG Text Sources};

\draw[decorate, decoration={brace, amplitude=5pt}, thick, purple!60] 
    (14.6, 1.5) -- (14.6, 9.3);
\node[right=0.2cm of classification, rotate=-90, font=\small, text=purple!80] at (15, 5.5) {Applications};

\node[text width=2.5cm, font=\tiny, text=gray!170] at (0, 2) {\textit{Examples:}\\• Climate risk reports\\• Labor practices\\• Governance compliance};

\node[text width=2.5cm, font=\tiny, text=gray!170] at (14, 1) {\textit{Examples:}\\• Topic categorization\\• Company clustering\\• ESG Q\&A systems};

\end{tikzpicture}%
}
\caption{Embeddings Applications}
\label{fig:embeddings_apps}
\end{figure*}
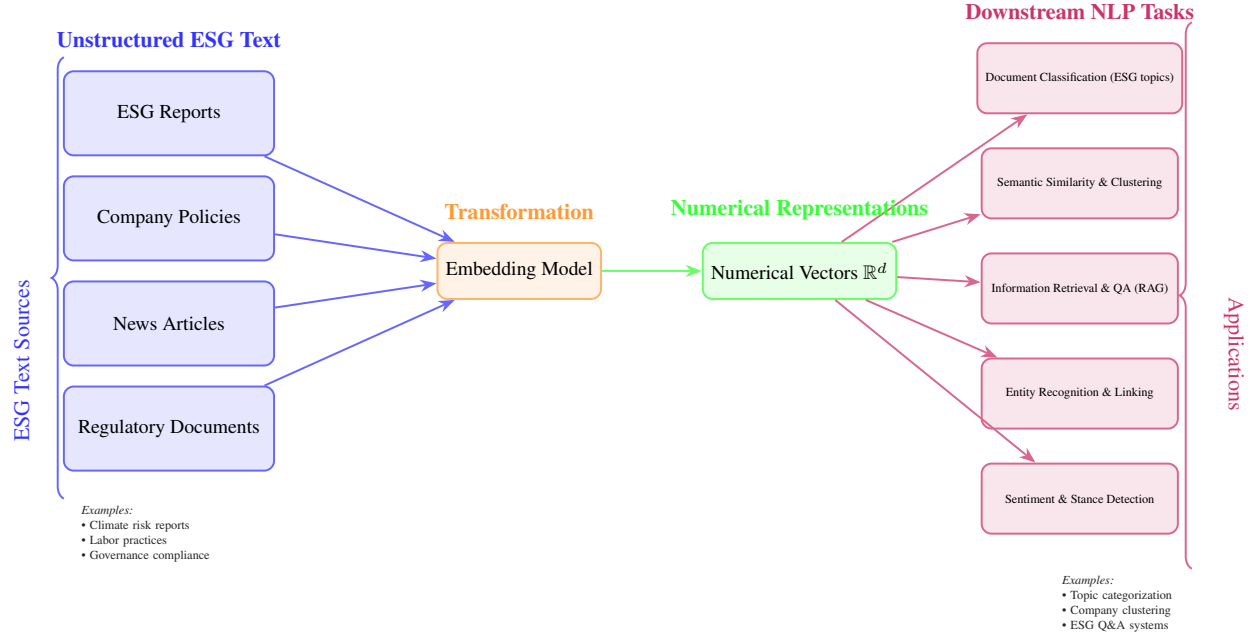

In this context, these models serve as a powerful approach for transforming unstructured ESG text into numerical representations suitable for downstream NLP tasks, see Figure \ref{fig:embeddings_apps}. In this domain, embeddings play a central role in:

\begin{itemize}
    \item Document classification (e.g., categorizing ESG reports by topic such as climate risk, labor practices, or governance compliance).
    \item Semantic similarity and clustering (e.g., grouping companies or policies with related ESG themes).
    \item Information retrieval and question answering (e.g., enabling retrieval-augmented generation, or RAG, systems that can ground ESG queries in large corpora of sustainability documents).
    \item Entity recognition and linking (e.g., identifying and disambiguating entities like organizations, emission targets, or regulatory frameworks across multiple sources).
    \item Sentiment and stance detection (e.g., measuring public or media sentiment around ESG controversies).
\end{itemize}

\begin{figure}[ht]
\centering
\begin{tikzpicture}[
  node distance=1.1cm and 0.8cm,
  box/.style={rectangle, draw=blue!60, fill=blue!10, thick, minimum height=0.7cm, minimum width=2.0cm, text centered},
  data/.style={rectangle, draw=orange!70, fill=orange!10, thick, minimum height=0.5cm, minimum width=2cm, text centered},
  arrow/.style={thick, ->, >=Stealth}
  ]
\node[box] (query) {User Query};
\node[box, right=of query] (encoder) {Query Encoder};
\node[data, above=of encoder] (corpus) {Document Corpus};
\node[box, right=of encoder] (matcher) {Semantic Matcher};
\node[box, below=of matcher] (ranker) {Ranker};
\node[box, left=of ranker] (results) {Ranked Results};

\draw[arrow] (query) -- (encoder);
\draw[arrow] (encoder) -- (matcher);
\draw[arrow] (corpus) -- (matcher) node[midway, right, font=\tiny] {Doc Embeddings};
\draw[arrow] (matcher) -- (ranker);
\draw[arrow] (ranker) -- (results);
\end{tikzpicture}
\caption{Information Retrieval IR system}
\label{fig:IR}
\end{figure}
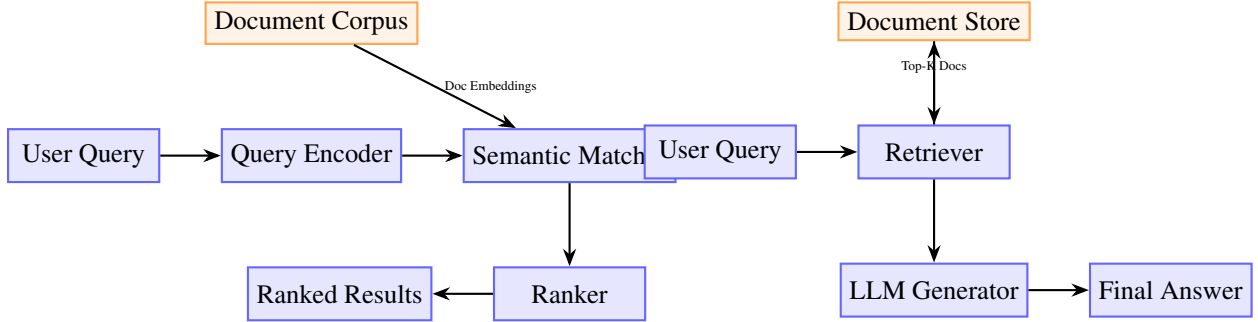

Information Retrieval (IR) is a fundamental task in natural language processing that focuses on finding and ranking relevant documents from a large corpus in response to a user's information need. Unlike traditional keyword-based search, modern IR systems leverage semantic understanding to match queries with documents based on meaning rather than exact lexical overlap. The effectiveness of IR systems is critical for numerous downstream applications, including question answering, document summarization, and as a foundational component in RAG architectures \cite{NLPBook25}. Figure \ref{fig:IR} illustrates the architecture of a typical neural IR system.

The mechanism of modern IR systems operates as follows:

\begin{enumerate}
    \item \textbf{Query Encoding:} Given a user query \( q \), an embedding model encodes it into a dense vector representation:
    \[
    \mathbf{v}_q = \text{Encoder}(q)
    \]
    where \( \mathbf{v}_q \in \mathbb{R}^d \) is the query embedding in a \( d \)-dimensional semantic space.
    
    \item \textbf{Documents Encoding:} Similarly, all documents \( \{d_i\}_{i=1}^N \) in the corpus are pre-encoded into dense vectors:
    \[
    \mathbf{v}_{d_i} = \text{Encoder}(d_i), \quad i = 1, \ldots, N
    \]
    These document embeddings are typically computed offline and stored in a vector database for efficient retrieval.
    
    \item \textbf{Similarity Computation:} Relevance scores between the query and each document are computed using a similarity function, typically cosine similarity:
    \[
    \text{sim}(q, d_i) = \frac{\mathbf{v}_q \cdot \mathbf{v}_{d_i}}{\|\mathbf{v}_q\| \|\mathbf{v}_{d_i}\|}
    \]
    
    \item \textbf{Ranking:} Documents are ranked by their similarity scores, and the top-\( k \) most relevant documents are returned:
    \[
    \{d_i\}_{i=1}^k = \text{top-}k\left(\text{argsort}_{d \in D}\{\text{sim}(q, d)\}\right)
    \]
\end{enumerate}

The quality of embedding models is crucial for IR performance, as they must effectively capture semantic relationships and domain-specific concepts. In specialized domains like ESG, where documents contain technical terminology, regulatory language, and domain-specific context, the choice of embedding model can significantly impact retrieval accuracy and downstream task performance.

Retrieval-Augmented Generation (RAG) is a cutting-edge architecture in NLP that fuses document retrieval with language generation, addressing a key limitation of large language models (LLMs): their inability to access external, up-to-date knowledge at inference time. Originally proposed by Facebook AI Research, RAG systems leverage real-time access to external corpora, enabling models to generate more factually accurate, context-aware answers \cite{NLPBook25}. Figure \ref{fig:RAG} illustrates the key components of a typical RAG system, which include: 

\begin{itemize}
    \item Retriever: Selects relevant documents. The retrieval quality directly impacts generation accuracy. 

    \item Generator: A pretrained LLM that synthesizes an output based on the retrieved documents plus the query.
\end{itemize}

\begin{figure}[ht]
\centering
\begin{tikzpicture}[
  node distance=1.1cm and 0.8cm,
  box/.style={rectangle, draw=blue!60, fill=blue!10, thick, minimum height=0.7cm, minimum width=2.0cm, text centered},
  data/.style={rectangle, draw=orange!70, fill=orange!10, thick, minimum height=0.5cm, minimum width=2cm, text centered},
  arrow/.style={thick, ->, >=Stealth}
  ]

\node[box] (user) {User Query};
\node[box, right=of user] (retriever) {Retriever};
\node[data, above=of retriever] (docstore) {Document Store};

\node[box, below=of retriever] (generator) {LLM Generator};
\node[box, right=of generator] (output) {Final Answer};

\draw[arrow] (user) -- (retriever);
\draw[arrow] (retriever) -- (docstore);
\draw[arrow] (docstore) -- (retriever) node[midway, above ] {\tiny Top-K Docs};
\draw[arrow] (retriever) -- (generator);
\draw[arrow] (generator) -- (output);

\end{tikzpicture}
\caption{RAG pipeline}
\label{fig:RAG}
\end{figure}

The mechanism of Retrieval-Augmented Generation (RAG) is as follows: 

\begin{enumerate}
    \item \textbf{User Query:} Let the input be a natural language query \( q \).
    
    \item \textbf{Retriever:} A dense retriever encodes the query into a vector and retrieves the top \( k \) documents from an external corpus \( D \). This can be formalized as:
    \[
    \{d_i\}_{i=1}^k = \text{Retriever}(q, D)
    \]
    where each \( d_i \) is a document retrieved based on semantic similarity.
    
    \item \textbf{Generator:} For each retrieved document \( d_i \), the generator model produces a response conditioned on the pair \( (q, d_i) \):
    \[
    y_i = \text{Generator}(q, d_i)
    \]
\end{enumerate}

This architecture enables dynamic knowledge integration without retraining the generator, as updating the external data is sufficient to refresh the model's knowledge.


While embeddings are widely used in NLP, their effectiveness in these ESG-specific tasks has not been systematically studied. ESG texts often contain domain-specific terminology, cross-lingual variations, and nuanced expressions of responsibility and risk, raising questions about how well general-purpose embedding models capture these subtleties.

In this paper, we focus on ESG embeddings for Retrieval-Augmented Generation (RAG) and Information Retrieval (IR) tasks. Despite the growing importance of ESG data in finance and sustainability reporting, there is currently no standardized benchmark to systematically evaluate how well embedding models perform on ESG-specific tasks. Existing general-purpose embedding benchmarks fail to capture the unique linguistic characteristics, specialized terminology, and conceptual complexity inherent in ESG discourse. To address this gap, we construct a benchmark dataset specifically tailored to the ESG domain. The benchmark dataset is generated using Generative AI and subsequently reviewed one by one by ESG domain experts to ensure quality control and accuracy of the generated data.
Using this benchmark dataset, we evaluate fourteen embedding models, comprising both open-source and closed-source (proprietary) solutions, comparing their performance across representative ESG tasks including semantic search, retrieval, and RAG-based question answering. It is important to note that our dataset is designed exclusively for benchmarking and evaluation purposes, not for fine-tuning or training embedding models. This approach enables a fair and systematic evaluation of how different embedding approaches handle the unique linguistic and conceptual challenges posed by ESG data, providing practitioners and researchers with actionable insights for model selection in real-world ESG applications.

The rest of this paper is organized as follows: Section \ref{sec:related} reviews related work, Section \ref{sec:data} describes the benchmark dataset creation process, Section \ref{sec:metrics} describes evaluation metrics, Section \ref{sec:models} presents the models considered for benchmarking, Section \ref{sec:results} presents benchmarking results, and finally Section \ref{sec:conclusion} draws conclusions.

\section{Related Work}
\label{sec:related}

\subsection{Embedding Model Evaluation and Benchmarking}
The systematic evaluation of text embedding models has become increasingly important as these models serve as fundamental components in various NLP pipelines. \cite{reimers2019sentence} introduced the Sentence-BERT framework and established early benchmarking practices for sentence embeddings using tasks such as semantic textual similarity and natural language inference. The Massive Text Embedding Benchmark (MTEB) by \cite{muennighoff2022mteb} significantly advanced this field by providing a comprehensive evaluation framework across multiple languages and domains, covering tasks including classification, clustering, pair classification, reranking, retrieval, semantic textual similarity, and summarization.
\cite{wang2022text} extended embedding evaluation to domain-specific contexts, demonstrating that model performance varies significantly across different domains and highlighting the need for specialized benchmarks. The work by \cite{thakur2021beir} on BEIR (Benchmarking Information Retrieval) specifically addressed the evaluation of dense retrieval systems, showing that models optimized for general tasks often underperform on specialized domains. More recently, \cite{xiao2023cmteb} introduced C-MTEB, a Chinese text embedding benchmark, emphasizing the importance of language-specific evaluation frameworks.

\subsection{Retrieval-Augmented Generation in Specialized Domains}
The application of Retrieval-Augmented Generation (RAG) systems to specialized domains has emerged as a promising approach for handling domain-specific knowledge bases. \cite{lewis2020retrieval} introduced the RAG framework, demonstrating its effectiveness in combining parametric and non-parametric knowledge. \cite{guu2020retrieval} explored similar concepts with REALM, showing improvements in knowledge-intensive NLP tasks.
In the financial and ESG context, recent work by \cite{ni2023finqa} explored the application of RAG systems to financial question answering, highlighting the importance of high-quality retrieval components for domain-specific applications. The work by \cite{izacard2021leveraging} on Fusion-in-Decoder further demonstrated how retrieval quality directly impacts generation performance, emphasizing the critical role of embedding models in RAG pipelines.

\subsection{ESG Data Characteristics and Challenges}
ESG data presents unique linguistic and structural challenges that distinguish it from general text corpora. \cite{christensen2022transparency} analyzed the heterogeneity of ESG reporting standards and highlighted the inconsistency in terminology and metrics across different frameworks such as GRI, SASB, and TCFD. This heterogeneity poses significant challenges for NLP models, as noted by \cite{bingler2022cheap}, who found that standard text classification approaches struggle with the ambiguity and context-dependency inherent in ESG disclosures.
The temporal evolution of ESG language has been examined by \cite{sautner2023natural}, who showed that ESG-related terminology and concepts evolve rapidly, requiring models to adapt to changing linguistic patterns. Furthermore, \cite{hoberg2017text} demonstrated the importance of document structure and context in financial text analysis, insights that are particularly relevant for ESG documents that often contain complex nested information and cross-references.

\subsection{Domain-Specific Text Embedding Applications}

The development of specialized embedding models for specific domains has consistently demonstrated advantages over general-purpose approaches across various fields. In the biomedical domain, BioBERT \cite{lee2020biobert} and ClinicalBERT \cite{huang2019clinicalbert} have shown substantial improvements over general BERT models for medical text analysis, effectively capturing domain-specific terminology and semantic relationships. Similarly, the financial domain has benefited from domain-specific pre-training, with \cite{araci2019finbert} and \cite{yang2020finbert} demonstrating significant performance improvements on financial text understanding tasks. These successes across diverse domains underscore the value of tailoring embedding models to capture the unique linguistic characteristics and conceptual structures of specialized fields.

The application of natural language processing techniques to Environmental, Social, and Governance (ESG) data represents a natural extension of this domain-specialization paradigm. Early foundational work by \cite{loughran2011liability} established the importance of domain-specific approaches for financial text analysis by developing specialized sentiment lexicons, highlighting that generic NLP tools often fail to capture the nuanced language of financial and sustainability-related documents. This insight proved particularly relevant for ESG analysis, as \cite{serafeim2022stock} demonstrated that standard NLP models struggle with the specialized terminology and contextual complexity prevalent in ESG disclosures and sustainability reporting.

Recent advances in ESG-specific NLP have built upon these foundations to develop increasingly sophisticated models. \cite{araci2019finbert} introduced FinBERT, a BERT model fine-tuned for financial text that showed marked improvements in financial sentiment analysis compared to general-purpose models. Extending this approach to the ESG domain specifically, \cite{huang2023esgbert} developed ESG-BERT, a transformer model trained on ESG-related documents that demonstrated superior performance on ESG classification tasks. In the environmental sustainability space, \cite{stammbach2022climatbert} developed ClimateBERT for climate change-related text classification, while \cite{varini2020goldfinch} focused on environmental policy document analysis. Most recently, \cite{kok2023climabert} proposed ClimaBERT, which targets climate-related financial disclosures and achieves notable improvements in climate risk assessment tasks. Collectively, these works establish a clear trajectory toward increasingly specialized models that can effectively handle the unique linguistic and conceptual challenges inherent in ESG data.

\subsection{Gap Analysis and Motivation}
Despite the growing body of work on ESG NLP applications and embedding model evaluation, several critical gaps remain. First, existing embedding benchmarks like MTEB, while comprehensive, lack domain-specific evaluation for ESG tasks and do not capture the unique challenges posed by sustainability-related text. 

Second, current ESG-focused NLP studies typically evaluate models on single tasks or datasets, limiting the generalizability of findings across the diverse ESG landscape. In other words, classification tasks have been studied, but IR/RAG are yet to be investigated. 

Third, the comparative evaluation of open-source versus proprietary embedding models in the ESG domain remains unexplored, despite the practical importance of this distinction for organizations with different resource constraints and data privacy requirements.

This work addresses these gaps by providing the first benchmark specifically designed for ESG embedding model evaluation, encompassing IR and RAG tasks, while comparing both open-source and proprietary approaches across multiple evaluation dimensions.


\section{Benchmark Dataset Creation}
\label{sec:data}

This section describes the methodology for constructing a benchmark dataset for search relevance in the ESG domain. The dataset is designed to evaluate embedding models across diverse ESG-related tasks such as document retrieval, semantic similarity, and retrieval-augmented generation (RAG).

\subsection{Dataset Structure}
The dataset follows a query–document–relevance triplet structure. Each entry consists of:
\begin{itemize}
    \item \textbf{Query:} The search term or question.
    \item \textbf{Documents:} A list of textual passages or articles.
    \item \textbf{Relevance Score:} A numerical label (e.g., 0--3) indicating how relevant a document is to the query.
\end{itemize}

\subsection{Query Generation}
We use OpenAI's ChatGPT-4.1 (GPT-4.1) as our generative model for creating diverse ESG-related queries. The use of generative AI for benchmark dataset creation offers several critical advantages over manual construction: 

\begin{enumerate}
    \item It enables rapid generation of diverse queries and documents at scale.
    \item It ensures systematic coverage across the three ESG pillars (Environmental, Social, and Governance) with controlled distribution.
    \item It allows for consistent structural formatting across all records.
\end{enumerate}

While generative AI provides the foundation for dataset creation, the subsequent expert validation process ensures that the benefits of automated generation are complemented by human domain expertise and quality control.

To ensure the quality and diversity of generated queries, we employ systematic prompt engineering techniques. We iteratively developed and refined multiple prompt templates, testing various formulations with different levels of specificity, context, and structural constraints. Through this iterative process, we evaluated the outputs based on criteria such as relevance to ESG topics, linguistic diversity, query complexity, and alignment with our aim. After extensive experimentation with numerous prompt variations, we identified the optimal prompts that consistently produced high-quality, diverse queries spanning the three ESG categories: \textit{Environmental}, \textit{Social}, and \textit{Governance}. Each generated query was subsequently reviewed by ESG domain experts to validate its appropriateness and ensure it accurately represents the types of information retrieval tasks encountered in practical ESG applications.

\noindent\textbf{Generic Prompt Example:}
\begin{lstlisting}
Generate N search queries related to <ESG category>.
\end{lstlisting}

\noindent\textbf{Example Output:}
\begin{lstlisting}
1. What are the company's carbon emission targets?
2. How does the company support renewable energy initiatives?
3. What policies are in place for employee well-being?
4. How does the company address data privacy concerns?
5. Is the company aligned with ESG reporting standards?
\end{lstlisting}

This process is repeated for all ESG categories to ensure balanced domain coverage.

\subsection{Document Generation}
For each query, ChatGPT is prompted to generate multiple candidate documents with varying levels of relevance.  

\noindent\textbf{Generic Prompt Example:}
\begin{lstlisting}
For the query "<Q>", generate three documents:
- One highly relevant document (relevance = 3).
- One moderately relevant document (relevance = 2).
- One irrelevant document (relevance = 0).

Return the output in valid JSON format.
\end{lstlisting}

\noindent\textbf{Example Output:}
\begin{lstlisting}
{
  "query": "What are the company's carbon emission targets?",
  "documents": [
    {"text": "The company aims for net-zero carbon emissions by 2030.",
     "relevance": 3, "label": "highly relevant"},
    {"text": "Investing in renewable energy is a company goal.",
     "relevance": 2, "label": "moderately relevant"},
    {"text": "The company reported record profits this year.",
     "relevance": 0, "label": "irrelevant"}
  ]
}
\end{lstlisting}

We generated benchmark data in both English and Italian. After expert review and elimination of low-quality records, we finalized the dataset with 624 records: 312 records in English and 312 records in Italian. This sample size is statistically sufficient and methodologically appropriate for embedding model evaluation in specialized domains.

The inclusion of Italian language data alongside English is motivated by this work being part of a broader Italian research project aimed at assessing and advancing language resources for Italian and other Latin languages. Italian, despite being a major European language with significant economic and regulatory importance in the ESG domain, remains underrepresented in NLP benchmarks compared to English. By creating a parallel Italian benchmark, we address this gap and enable fair evaluation of embedding models' multilingual capabilities, which is crucial for organizations operating across European markets where ESG reporting requirements span multiple languages.

Each record contains a query and three associated documents with graded relevance scores, designed to evaluate the embedding models' ability to distinguish between highly relevant, partially relevant, and irrelevant content in ESG contexts.

\subsection{Dataset Balancing and Quality Assurance}
To ensure fairness, the dataset is balanced across ESG categories (\textit{Environmental, Social, Governance}) and relevance levels (highly relevant, moderately relevant, and irrelevant). This prevents over-representation of specific topics or labels.

To guarantee quality and domain validity, we implemented a rigorous quality assurance (QA) process. All 624 records were reviewed individually by an ESG domain expert with extensive experience in sustainability reporting and corporate governance. The expert evaluation focused on multiple dimensions: 

\begin{enumerate}
    \item Semantic accuracy of the query with respect to ESG terminology and concepts.
    \item Appropriateness of the relevance grades assigned to each document.
    \item Factual correctness of the content in relation to ESG standards and frameworks.
    \item Alignment with real-world ESG information retrieval scenarios.
\end{enumerate}

During this review process, records exhibiting ambiguity in relevance labeling, factual inaccuracies, inconsistent terminology, or poor query-document alignment were flagged for revision. Records that violated one or more of the quality assurance dimensions and could not be sufficiently corrected were excluded from the final dataset. Through this rigorous filtering process, 63 records per language were eliminated, resulting in a final benchmark dataset of 312 records in English and 312 records in Italian (624 records total). This iterative review and refinement process ensured that the final benchmark dataset maintains high consistency, accuracy, and representativeness of genuine ESG information retrieval tasks. The expert-validated nature of our dataset distinguishes it from purely automated benchmarks and provides confidence in its reliability for evaluating embedding model performance in the ESG domain.

\noindent\textbf{Generic Prompt Example for Large-Scale Generation:}
\begin{lstlisting}
Task:
Generate N search queries in <language> related to the Environmental, Social, and Governance (ESG) domain. Each query should belong to one of the following categories:

Environmental (E): Carbon footprint, renewable energy, climate risk, waste management.

Social (S): Diversity and inclusion, employee well-being, corporate social responsibility (CSR).

Governance (G): Corporate ethics, executive compensation, board diversity, regulatory compliance.

Ensure balanced queries across domains.

For each query, generate three documents in <language>:
- Relevant (relevance = 3).
- Somewhat relevant (relevance = 2).
- Irrelevant (relevance = 0).

Return output in valid JSON format including the domain.
\end{lstlisting}

\subsection{Workflow Illustration}
\begin{figure}[ht!]
\centering
\begin{tikzpicture}[
    node distance=1.2cm, 
    auto, 
    >=latex,
    process/.style={
        rectangle, 
        rounded corners=2pt, 
        draw=black, 
        fill=blue!15,
        text centered, 
        minimum height=0.7cm, 
        minimum width=2.2cm,
        font=\scriptsize
    },
    arrow/.style={->, thick, color=blue!70}
]
\node[process] (query) {Query Generation};
\node[process, below of=query] (docs) {Document Generation};
\node[process, below of=docs] (label) {Relevance Labeling};
\node[process, below of=label] (balance) {Dataset Balancing};
\node[process, below of=balance] (qa) {Expert Review and QA};
\draw[arrow] (query) -- (docs);
\draw[arrow] (docs) -- (label);
\draw[arrow] (label) -- (balance);
\draw[arrow] (balance) -- (qa);
\end{tikzpicture}
\caption{Workflow for ESG benchmark dataset creation: from query generation to expert validation.}
\label{fig:dataset_workflow}
\end{figure}
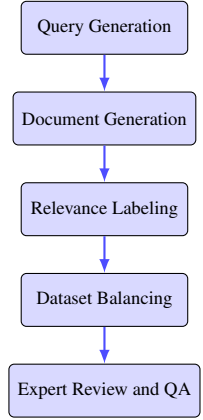

Figure~\ref{fig:dataset_workflow} illustrates the dataset creation workflow, from query generation to expert validation.






\section{Evaluation Metrics}
\label{sec:metrics}

To objectively measure the performance of embedding models in information retrieval tasks, we evaluate the quality of search results in a Retrieval-Augmented Generation (RAG) pipeline using standard metrics: Precision@K and Normalized Discounted Cumulative Gain (NDCG@K). 
All of these metrics are evaluated at cutoff $K$, meaning we only consider the top $K$ retrieved results. 
Each metric takes values between 0 and 1, where higher values indicate better performance. 

Broadly, evaluation metrics can be divided into two groups:  
\begin{itemize}
    \item \textbf{Not rank-aware}: metrics such as Precision, which only consider the presence of relevant items in the top $K$ results, regardless of their position.  
    \item \textbf{Rank-aware}: metrics such as NDCG, which account for both the number of relevant items and their rank position in the retrieved list.  
\end{itemize}

In this section, we formally define the metrics used in our benchmarking study.

\subsection{Precision@K}
The Precision@K metric measures the proportion of retrieved items in the top $K$ that are relevant. It is not rank-aware and does not account for the ordering of results.

\begin{equation}
\text{Precision@K} = \frac{TP}{TP + FP} = \frac{TP}{K}
\label{eq:precision_at_k}
\end{equation}

where $TP$ is the number of true positives and $FP$ is the number of false positives within the top $K$ results.  












\subsection{Normalized Discounted Cumulative Gain (NDCG@K)}
Normalized Discounted Cumulative Gain@K (NDCG@K) measures how well a system ranks items by relevance. Unlike binary relevance metrics, NDCG supports graded relevance levels (e.g., highly relevant, moderately relevant, irrelevant).  

It is defined as:

\begin{equation}
\text{NDCG@K} = \frac{\text{DCG@K}}{\text{IDCG@K}}
\label{eq:ndcg}
\end{equation}

where DCG is the Discounted Cumulative Gain:

\begin{equation}
\text{DCG@K} = \sum_{i=1}^{K} \frac{\text{rel}_i}{\log_2(i + 1)}
\label{eq:dcg_at_k}
\end{equation}

and IDCG is the maximum possible DCG score (ideal ranking).

Together, these metrics provide a comprehensive view of retrieval quality in ESG-focused RAG pipelines. Precision highlights coverage of relevant information, while NDCG assesses whether embedding models rank the most critical ESG documents at the top of the list.

\subsection{Evaluation Algorithm}

This code evaluates information retrieval (IR) systems by comparing different embedding models. The code benchmarks multiple embedding models on their ability to match queries with relevant documents using semantic similarity.

\begin{algorithm}
\caption{Evaluate Multiple Embedding Models on IR Dataset}
\begin{algorithmic}[1]
\Require Dataset $D$ with queries and documents
\Require Set of embedding models $M = \{m_1, m_2, \ldots, m_k\}$
\Ensure $D$ updated with NDCG and precision scores for each model

\State \textbf{CosineSimilarity}($\vec{v_1}, \vec{v_2}$) $= \frac{\vec{v_1} \cdot \vec{v_2}}{\|\vec{v_1}\| \times \|\vec{v_2}\|}$

\For{each row $i$ in $D$}
    \State $query \gets D[i].query$
    \State $documents \gets D[i].documents$
    \State $relevanceScores \gets [doc.relevance \mid doc \in documents]$
    \State $relevanceLabels \gets [doc.label \mid doc \in documents]$
    
    \For{each $model \in M$}
        \State $queryEmbedding \gets D[i].query\_model$
        \State $docEmbeddings \gets D[i].documents\_model$
        
        \State \textit{// Compute similarities between query and all documents}
        \State $similarities \gets []$
        \For{each $docEmb$ in $docEmbeddings$}
            \State $sim \gets \textbf{CosineSimilarity}(queryEmbedding, docEmb)$
            \State Append $sim$ to $similarities$
        \EndFor
        
        \State \textit{// Compute evaluation metrics}
        \State $ndcg \gets$ compute NDCG from $relevanceScores$ and $similarities$
        
        \State \textit{// Binarize scores and similarities for precision}
        \State $binaryRelevance \gets [1 \text{ if } score > 1 \text{ else } 0 \mid score \in relevanceScores]$
        \State $binaryPredictions \gets [1 \text{ if } sim > 0.5 \text{ else } 0 \mid sim \in similarities]$
        \State $precision \gets$ compute precision from $binaryRelevance$ and $binaryPredictions$
        
        \State \textit{// Store results in dataframe}
        \State $D[i].model\_ndcg \gets ndcg$
        \State $D[i].model\_precision \gets precision$
    \EndFor
\EndFor
\end{algorithmic}
\end{algorithm}

\section{Embedding Models Considered}
\label{sec:models}

To evaluate embedding performance in the ESG domain, we benchmark both open-source and closed-source models.
Open-source models are deployed locally via the \texttt{Ollama} framework, while closed-source models are accessed through the OpenAI and Google Gemini APIs.
This setup enables us to compare transparent, community-driven models with proprietary, large-scale industry models.

\subsection{Open-Source Models (Ollama)}
The following open-source embedding models were included in our benchmark experiments:

\begin{itemize}
    \item \textbf{nomic-embed-text}: A general-purpose embedding model optimized for text understanding and retrieval tasks, designed for scalability and efficiency in vector search applications \cite{nussbaum2024nomic}.
    \item \textbf{nomic-embed-text-v2-moe}: An updated Mixture-of-Experts variant of the Nomic embedding model, offering improved representation quality over its predecessor \cite{nussbaum2024nomic}.
    \item \textbf{mxbai-embed-large}: A large embedding model emphasizing semantic representation quality, suitable for tasks requiring high retrieval accuracy and nuanced contextual understanding \cite{mxbai1,mxbai2}.
    \item \textbf{bge-m3}: A multilingual embedding model from the \texttt{BAAI General Embedding (BGE)} family, supporting cross-lingual retrieval and particularly relevant for multilingual ESG data \cite{chen2024bge}.
    \item \textbf{all-MiniLM}: A lightweight embedding model based on the MiniLM architecture, providing fast inference with competitive performance for resource-constrained environments \cite{wang2020minilm}.
    \item \textbf{embeddinggemma:300m}: A compact 300M-parameter embedding model from Google’s Gemma family, designed for efficient semantic retrieval tasks \cite{gemma2024embedding}.
    \item \textbf{qwen3-embedding:0.6b}: A 0.6B-parameter embedding model from Alibaba’s Qwen3 series, offering strong multilingual and domain-specific retrieval capabilities \cite{qwen3embedding2025}.
    \item \textbf{qwen3-embedding:4b}: A larger 4B-parameter variant of the Qwen3 embedding series, providing higher-capacity semantic representations for complex retrieval tasks \cite{qwen3embedding2025}.
    \item \textbf{granite-embedding:278m}: A 278M-parameter embedding model from IBM’s Granite series, tailored for enterprise and domain-specific text retrieval \cite{granite2024embedding}.
\end{itemize}

\subsection{Closed-Source Models (OpenAI)}
In addition to open-source models, we benchmark widely used embedding models provided by OpenAI:

\begin{itemize}
    \item \textbf{text-embedding-3-small}: A cost-efficient embedding model optimized for semantic similarity and retrieval tasks.
    \item \textbf{text-embedding-3-large}: A larger and more powerful variant designed for higher accuracy in tasks requiring fine-grained semantic representation.
    \item \textbf{text-embedding-ada-002}: One of OpenAI’s earlier widely adopted embedding models, known for its balance between performance and efficiency.
\end{itemize}

\subsection{Closed-Source Models (Google Gemini)}
We include two models from Google's Gemini embedding API:

\begin{itemize}
    \item \textbf{gemini-embedding-001}: Google's text embedding model accessible via the Gemini API, offering strong performance on semantic retrieval and similarity tasks \cite{google2024textembedding004}.
    \item \textbf{gemini-embedding-2}: Google's native multimodal embedding model that maps text, images, video, audio, and documents into a unified embedding space, achieving state-of-the-art performance on a variety of embedding benchmarks \cite{shanbhogue2026gemini}.
\end{itemize}

Table \ref{tab:embedding_models} summarizes all embedding models evaluated in our benchmark, detailing their providers, dimensions, and brief descriptions.

\begin{table*}[ht]
\centering
\caption{Embedding models evaluated in our benchmark, along with their providers, dimensions, and citations.}
\resizebox{\textwidth}{!}{%
\begin{tabular}{llrlc}
\hline
\textbf{Model Name} & \textbf{Provider} & \textbf{Dimension} & \textbf{Description} & \textbf{Citation} \\
\midrule
nomic-embed-text        & Nomic AI       & 768  & General-purpose open-source retrieval model        & \cite{nussbaum2024nomic} \\
nomic-embed-text-v2-moe & Nomic AI       & 768  & Mixture-of-Experts variant of nomic-embed-text     & \cite{nussbaum2024nomic} \\
mxbai-embed-large       & Mixedbread AI  & 1024 & High-performance model optimized for retrieval     & \cite{mxbai1,mxbai2} \\
bge-m3                  & BAAI           & 1024 & Multilingual model supporting long inputs          & \cite{chen2024bge} \\
all-MiniLM              & Microsoft      & 384  & Lightweight model for efficient inference          & \cite{wang2020minilm} \\
embeddinggemma:300m     & Google         & 768  & Compact Gemma-based embedding model                & \cite{gemma2024embedding} \\
qwen3-embedding:0.6b    & Alibaba        & 1024 & Qwen3 embedding, 0.6B parameters                  & \cite{qwen3embedding2025} \\
qwen3-embedding:4b      & Alibaba        & 2560 & Qwen3 embedding, 4B parameters                    & \cite{qwen3embedding2025} \\
granite-embedding:278m  & IBM            & 768  & Enterprise-focused Granite embedding model         & \cite{granite2024embedding} \\
text-embedding-3-small  & OpenAI         & 1536 & Cost-effective OpenAI embedding model              & \cite{openai2024embeddings} \\
text-embedding-3-large  & OpenAI         & 3072 & High-capacity OpenAI embedding model               & \cite{openai2024embeddings} \\
text-embedding-ada-002  & OpenAI         & 1536 & OpenAI’s balanced embedding model                  & \cite{openai2022embeddings} \\
gemini-embedding-001    & Google Gemini  & 768  & Google’s text embedding model via Gemini API       & \cite{google2024textembedding004} \\
gemini-embedding-2      & Google Gemini  & 3072 & Google\u2019s native multimodal embedding model         & \cite{shanbhogue2026gemini} \\
\bottomrule
\end{tabular}%
}
\label{tab:embedding_models}
\end{table*}


The selected models span a range of sizes, architectures, and licensing schemes.  
By including both open-source and closed-source models, our benchmark captures trade-offs between transparency, reproducibility, performance, and cost.  
This comparison provides practical insights into which models are better suited for ESG-specific tasks such as semantic search, classification, and Retrieval-Augmented Generation (RAG).

\section{Benchmarking Results}
\label{sec:results}

We evaluated fourteen embedding models on the ESG benchmark dataset in both English and Italian. Tables~\ref{tab:eng_results} and~\ref{tab:ita_results} therefore report NDCG, Precision, MAP, and MRR, sorted by NDCG for English and Italian respectively. We omit Recall@K from the tables because, in the evaluation code, \texttt{recall\_at\_k} defaults to \texttt{k = len(similarities)} and is called without an explicit cutoff; as a result, each query is evaluated over its full candidate list, making Recall@K equal to 1.0 for all models and uninformative for comparison.

\begin{table*}[htbp]
\centering
\caption{Performance Results on English ESG Benchmark Dataset (sorted by NDCG)}
\label{tab:eng_results}
\resizebox{\textwidth}{!}{%
\begin{tabular}{llcccc}
\hline
\textbf{Model} & \textbf{Provider} & \textbf{NDCG} & \textbf{Precision} & \textbf{MAP} & \textbf{MRR} \\
\hline
qwen3-embedding-4b      & Alibaba        & \textbf{0.9830} & 0.9781          & 0.9989          & \textbf{1.0000} \\
qwen3-embedding-0.6b    & Alibaba        & 0.9784          & 0.9818          & 0.9995          & \textbf{1.0000} \\
gemini-embedding-2      & Google Gemini  & 0.9771          & 0.6799          & 0.9994          & \textbf{1.0000} \\
text-embedding-3-large  & OpenAI         & 0.9731          & 0.9776          & \textbf{1.0000} & \textbf{1.0000} \\
gemini-embedding-001    & Google Gemini  & 0.9727          & 0.7304          & \textbf{1.0000} & \textbf{1.0000} \\
mxbai-embed-large       & Mixedbread AI  & 0.9703          & 0.9327          & \textbf{1.0000} & \textbf{1.0000} \\
text-embedding-ada-002  & OpenAI         & 0.9667          & 0.6667          & \textbf{1.0000} & \textbf{1.0000} \\
granite-embedding-278m  & IBM            & 0.9646          & 0.7671          & 0.9949          & 0.9984          \\
text-embedding-3-small  & OpenAI         & 0.9643          & \textbf{0.9840} & 0.9995          & \textbf{1.0000} \\
bge-m3                  & BAAI           & 0.9619          & 0.9071          & 0.9989          & \textbf{1.0000} \\
embeddinggemma-300m     & Google         & 0.9617          & 0.9829          & \textbf{1.0000} & \textbf{1.0000} \\
all-minilm              & Microsoft      & 0.9584          & \textbf{0.9840} & 0.9976          & 0.9984          \\
nomic-embed-text        & Nomic AI       & 0.9539          & 0.8419          & 0.9984          & \textbf{1.0000} \\
nomic-embed-text-v2-moe & Nomic AI       & 0.9518          & 0.9829          & 0.9981          & 0.9984          \\
\hline
\end{tabular}%
}
\end{table*}

\begin{table*}[htbp]
\centering
\caption{Performance Results on Italian ESG Benchmark Dataset (sorted by NDCG)}
\label{tab:ita_results}
\resizebox{\textwidth}{!}{%
\begin{tabular}{llcccc}
\hline
\textbf{Model} & \textbf{Provider} & \textbf{NDCG} & \textbf{Precision} & \textbf{MAP} & \textbf{MRR} \\
\hline
text-embedding-3-large  & OpenAI         & \textbf{0.9745} & \textbf{0.9861} & \textbf{1.0000} & \textbf{1.0000} \\
qwen3-embedding-0.6b    & Alibaba        & 0.9717          & 0.9514          & 0.9984          & \textbf{1.0000} \\
qwen3-embedding-4b      & Alibaba        & 0.9704          & 0.9765          & 0.9995          & \textbf{1.0000} \\
gemini-embedding-2      & Google Gemini  & 0.9686          & 0.6800          & 0.9994          & \textbf{1.0000} \\
granite-embedding-278m  & IBM            & 0.9647          & 0.7991          & 0.9979          & \textbf{1.0000} \\
embeddinggemma-300m     & Google         & 0.9643          & 0.9749          & 0.9995          & \textbf{1.0000} \\
bge-m3                  & BAAI           & 0.9634          & 0.9306          & 0.9995          & \textbf{1.0000} \\
text-embedding-3-small  & OpenAI         & 0.9611          & 0.9829          & \textbf{1.0000} & \textbf{1.0000} \\
text-embedding-ada-002  & OpenAI         & 0.9604          & 0.6667          & 0.9987          & 0.9984          \\
nomic-embed-text-v2-moe & Nomic AI       & 0.9511          & 0.9818          & 0.9995          & \textbf{1.0000} \\
mxbai-embed-large       & Mixedbread AI  & 0.9488          & 0.7099          & 0.9904          & \textbf{1.0000} \\
nomic-embed-text        & Nomic AI       & 0.9375          & 0.6667          & 0.9784          & 0.9952          \\
gemini-embedding-001    & Google Gemini  & 0.9286          & 0.6667          & \textbf{1.0000} & \textbf{1.0000} \\
all-minilm              & Microsoft      & 0.9160          & 0.8568          & 0.9439          & 0.9776          \\
\hline
\end{tabular}%
}
\end{table*}

\textbf{English Results --- Open-Source Models Dominate:} For English ESG content, the two top-performing models are open-source: \textbf{qwen3-embedding-4b} achieves the highest NDCG (0.983) and \textbf{qwen3-embedding-0.6b} ranks second (0.978), both outperforming all closed-source alternatives. \textbf{gemini-embedding-2} ranks third (NDCG 0.977), establishing itself as the strongest closed-source model for English ESG retrieval. This demonstrates that recent open-source embedding models have surpassed proprietary solutions on ESG-specific retrieval tasks.

\textbf{Italian Results --- Closed-Source Leads:} The ranking shifts for Italian. \textbf{text-embedding-3-large} achieves the highest NDCG (0.975) and the highest precision (0.986), outperforming all open-source models. \textbf{gemini-embedding-2} ranks fourth (NDCG 0.969), maintaining strong cross-lingual performance. This suggests that larger multilingual training corpora in closed-source models confer an advantage for non-English ESG content, where domain-specific Italian terminology is less represented in open-source pretraining data.

\textbf{Cross-Language Degradation:} Several models show notable NDCG drops from English to Italian. \textbf{all-minilm} drops from 0.958 to 0.916 and \textbf{mxbai-embed-large} from 0.970 to 0.949, suggesting limited multilingual generalization. By contrast, \textbf{text-embedding-3-large} improves slightly from 0.973 to 0.975, confirming its strong multilingual capability.

\textbf{Precision vs.\ Ranking Gaps:} \textbf{granite-embedding-278m} and \textbf{text-embedding-ada-002} consistently show competitive NDCG scores alongside low precision, indicating these models rank relevant documents well but struggle with binary retrieval thresholds. \textbf{gemini-embedding-2} and \textbf{gemini-embedding-001} exhibit the same pattern with precision below 0.74 in both languages despite strong NDCG. \textbf{text-embedding-ada-002} is included as a legacy baseline.

\textbf{MAP and MRR:} Most models achieve MAP and MRR near or equal to 1.0 in both languages, reflecting that the top-ranked document is almost always relevant. \textbf{all-minilm} shows the largest MAP drop in Italian (0.944), consistent with its NDCG degradation.

\textbf{Best Model Recommendations:} For English ESG RAG applications, \textbf{qwen3-embedding-4b} is the top choice (NDCG=0.983); \textbf{qwen3-embedding-0.6b} provides near-identical performance at 0.6B parameters. For Italian or multilingual ESG deployments, \textbf{text-embedding-3-large} is recommended. Among open-source models for Italian, \textbf{qwen3-embedding-0.6b} (NDCG=0.972) remains competitive. For applications requiring a unified multimodal embedding space, \textbf{gemini-embedding-2} offers competitive NDCG across both languages.

\subsection{Impact of Embedding Dimension on Retrieval Quality}
\label{sec:dim_experiment}

\begin{figure}[htbp]
\centering
\includegraphics[width=\columnwidth]{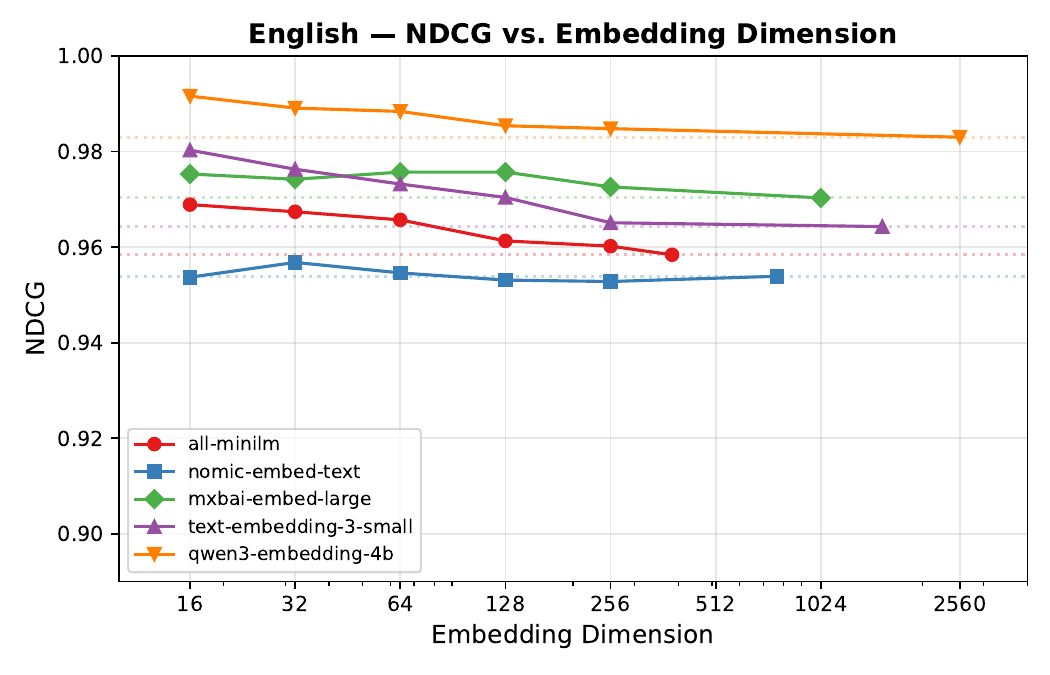}
\caption{NDCG vs.\ embedding dimension after PCA reduction (English)}
\label{fig:dim_eng}
\end{figure}

\begin{figure}[htbp]
\centering
\includegraphics[width=\columnwidth]{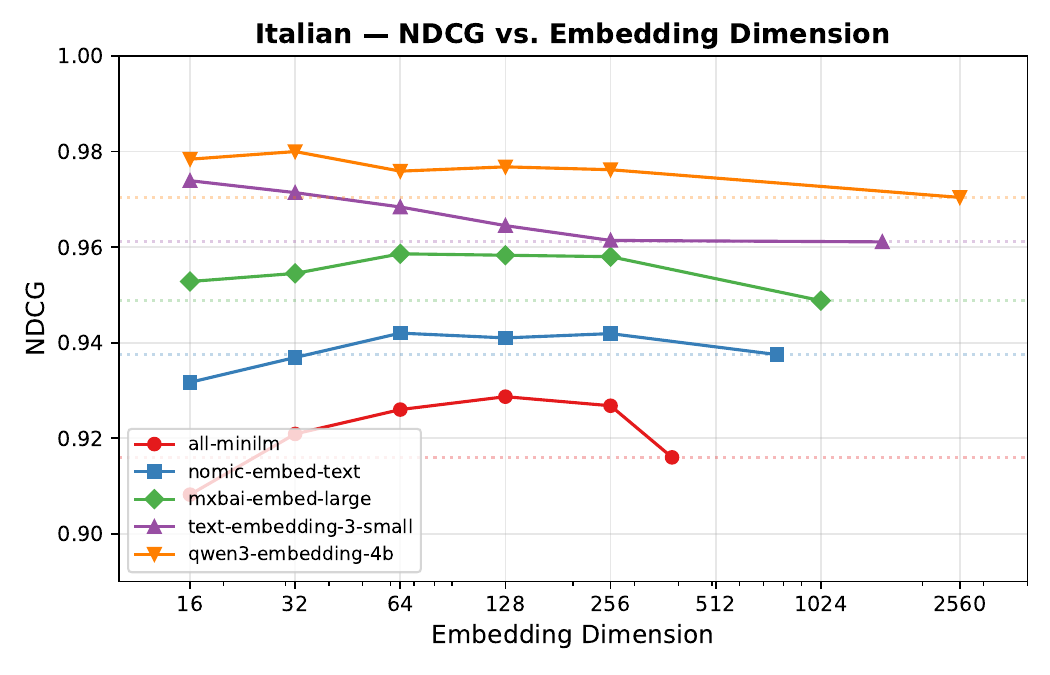}
\caption{NDCG vs.\ embedding dimension after PCA reduction (Italian)}
\label{fig:dim_ita}
\end{figure}

We further investigate the trade-off between embedding dimensionality and retrieval quality by applying Principal Component Analysis (PCA) to reduce the output dimensions of selected models and re-evaluating NDCG at each reduced dimension. Tables~\ref{tab:dim_eng} and~\ref{tab:dim_ita} report the numerical results, while Figures~\ref{fig:dim_eng} and~\ref{fig:dim_ita} visualize the trends for English and Italian, respectively.

\begin{table*}[htbp]
\centering
\caption{NDCG vs.\ embedding dimension after PCA reduction (English)}
\label{tab:dim_eng}
\resizebox{\textwidth}{!}{%
\begin{tabular}{lrrrrrrr}
\hline
\textbf{Model} & \textbf{Full Dim} & \textbf{NDCG@Full} & \textbf{NDCG@256} & \textbf{NDCG@128} & \textbf{NDCG@64} & \textbf{NDCG@32} & \textbf{NDCG@16} \\
\midrule
all-minilm              & 384  & 0.9584 & 0.9602 & 0.9613 & 0.9657 & 0.9674 & 0.9689 \\
nomic-embed-text        & 768  & 0.9539 & 0.9528 & 0.9531 & 0.9546 & 0.9568 & 0.9537 \\
mxbai-embed-large       & 1024 & 0.9703 & 0.9726 & 0.9757 & 0.9757 & 0.9742 & 0.9753 \\
text-embedding-3-small  & 1536 & 0.9643 & 0.9651 & 0.9704 & 0.9732 & 0.9763 & 0.9803 \\
qwen3-embedding-4b      & 2560 & 0.9830 & 0.9848 & 0.9854 & 0.9884 & 0.9891 & 0.9916 \\
\bottomrule
\end{tabular}%
}
\end{table*}

\begin{table*}[htbp]
\centering
\caption{NDCG vs.\ embedding dimension after PCA reduction (Italian)}
\label{tab:dim_ita}
\resizebox{\textwidth}{!}{%
\begin{tabular}{lrrrrrrr}
\hline
\textbf{Model} & \textbf{Full Dim} & \textbf{NDCG@Full} & \textbf{NDCG@256} & \textbf{NDCG@128} & \textbf{NDCG@64} & \textbf{NDCG@32} & \textbf{NDCG@16} \\
\midrule
all-minilm              & 384  & 0.9160 & 0.9268 & 0.9287 & 0.9260 & 0.9209 & 0.9082 \\
nomic-embed-text        & 768  & 0.9375 & 0.9419 & 0.9410 & 0.9420 & 0.9369 & 0.9317 \\
mxbai-embed-large       & 1024 & 0.9488 & 0.9580 & 0.9583 & 0.9586 & 0.9545 & 0.9528 \\
text-embedding-3-small  & 1536 & 0.9611 & 0.9614 & 0.9645 & 0.9684 & 0.9714 & 0.9739 \\
qwen3-embedding-4b      & 2560 & 0.9704 & 0.9762 & 0.9768 & 0.9759 & 0.9800 & 0.9784 \\
\bottomrule
\end{tabular}%
}
\end{table*}

The results reveal a notable finding: across nearly all models and both languages, reducing the embedding dimension via PCA does not degrade retrieval quality, even at aggressive reductions to 16 dimensions. In many cases, NDCG improves marginally after PCA reduction, suggesting that the leading principal components capture the most informative signal for retrieval while lower-variance dimensions contain noise. For example, \textbf{text-embedding-3-small} improves from 0.9643 to 0.9803 in English and from 0.9611 to 0.9739 in Italian when reduced to 16 dimensions. The \textbf{qwen3-embedding-4b} model maintains near-perfect NDCG across all reduced dimensions in both languages.

\section{Conclusion and Future Work}
\label{sec:conclusion}

This paper presents the first comprehensive benchmark for evaluating embedding models specifically tailored to Environmental, Social, and Governance (ESG) data analysis. Our study addresses a critical gap in the literature by providing systematic evaluation of both open-source and proprietary embedding models across representative ESG tasks, including classification, semantic similarity, retrieval, and Retrieval-Augmented Generation (RAG).

Our benchmark evaluation across fourteen embedding models reveals significant performance variations when applied to ESG-specific tasks. For English content, the open-source \textbf{qwen3-embedding-4b} achieves the highest NDCG (0.983), demonstrating that recent open-source models have surpassed proprietary solutions in English ESG retrieval. For Italian content, \textbf{text-embedding-3-large} leads with NDCG of 0.975 and the highest precision (0.986), highlighting the importance of strong multilingual training for non-English ESG deployments. Recall@K equals 1.0 for all models in both languages, while MAP and MRR are near-perfect across the board, confirming robust ranking behavior. Precision scores vary more widely (0.667--0.986), exposing meaningful differences in binary retrieval accuracy relevant for regulatory compliance tasks.

Our findings provide actionable insights for practitioners working with ESG data. For English ESG information retrieval and RAG applications, \textbf{qwen3-embedding-4b} is the top recommendation, with \textbf{qwen3-embedding-0.6b} offering near-equivalent performance at 0.6B parameters. For multilingual or Italian-language deployments, \textbf{text-embedding-3-large} is preferred. For organizations with constraints on proprietary model usage, \textbf{qwen3-embedding-0.6b} provides the best open-source option across both languages.

The quality assurance performed on the benchmark dataset ensures that our evaluation results are more representative of practical deployment scenarios.

This work makes several important contributions to both the ESG and NLP research communities. First, we establish the first standardized benchmark specifically designed for ESG embedding model evaluation, enabling fair and systematic comparison of different approaches. Second, our comprehensive evaluation across multiple task types provides insights into how different embedding models handle the unique linguistic and conceptual challenges posed by ESG data. Third, our inclusion of RAG-based evaluation represents a significant advance in understanding how embedding quality impacts downstream generative tasks in domain-specific contexts. This is particularly relevant given the growing adoption of RAG systems in corporate ESG workflows and sustainability reporting processes.

While our benchmark provides valuable insights, several limitations should be acknowledged. First, our evaluation focuses on English and Italian language ESG documents, and future work should extend the benchmark to multilingual ESG content to reflect the global nature of sustainability reporting. Second, we evaluated fourteen embedding models; while these represent a diverse range of architectures and approaches, the rapidly expanding landscape of embedding models means that many potentially relevant models were not included in our assessment. Third, the rapidly evolving landscape of embedding models necessitates regular updates to the benchmark to include newer architectures and approaches. Fourth, our benchmark dataset, while carefully constructed, may not fully capture the diversity of ESG documentation across all industries, company sizes, and reporting frameworks. Finally, we did not evaluate the computational costs, inference speed, or scalability of different models, which are practical considerations for real-world deployments. 

Future research directions include investigating the impact of domain-specific fine-tuning on embedding performance for ESG tasks, exploring the integration of multimodal data (text, numerical, and graphical information) commonly found in sustainability reports, and developing specialized evaluation metrics that better capture the nuanced requirements of ESG applications. The temporal evolution of ESG language and terminology also presents an interesting avenue for future investigation. As sustainability frameworks and reporting standards continue to evolve, understanding how embedding models adapt to changing linguistic patterns in ESG discourse will become increasingly important. Additionally, examining model performance across different ESG sub-domains (environmental vs. social vs. governance) and investigating the interpretability of embedding representations would provide valuable insights for practitioners.

As ESG considerations become increasingly central to corporate decision-making and regulatory compliance, the need for effective computational tools to process and analyze ESG data continues to grow. Our benchmark provides a foundation for systematic evaluation of embedding models in this critical domain, enabling researchers and practitioners to make informed decisions about model selection for their specific ESG applications. The significant performance variations we observed across different models underscore the importance of domain-specific evaluation rather than relying solely on general-purpose benchmarks. We hope that our work will encourage further research into ESG-specific NLP applications and contribute to the development of more effective tools for sustainability analysis and reporting. We make our benchmark dataset and evaluation code publicly available to facilitate reproducible research and encourage the broader research community to build upon our work. This open approach aligns with the transparency and accountability principles that are fundamental to the ESG domain itself.

\section*{Acknowledgments}
This research was partially supported by grant from  Italian Research Center on High Performance Computing, Big Data and Quantum Computing (ICSC) funded by EU\-NextGenerationEU (PNRR\-HPC,~CUP:C83C22000560007).

\bibliography{ref}

\end{document}